\documentclass[12pt]{article}
\usepackage{graphicx}
\input{epsf.tex}
\def\eqwithrates#1#2{\mathrel{\mathop{\rightleftharpoons}\limits
^{#1}_{#2}}}
\title{$NF$-$\kappa B$ oscillations and cell-to-cell variability}

\author{F. Hayot\footnote{Corresponding author. Tel: +1 614 292 2343; fax:
+1 614 292 7557. {\it email address}: hayot@mps.ohio-state.edu. Address
after September 1: Department of Neurology, Mount Sinai School of
Medicine, Box 1137, One Gustave L. Levy Place, New York, NY 10029} and C.
Jayaprakash\\ Department of Physics\\
Ohio State University\\
Columbus, OH 43210-1106, USA}

\begin{document}
\maketitle
\begin{abstract}
Oscillations in the transcriptional activator $NF$-$\kappa B$
localized in the nucleus have been observed when a cell is
stimulated by an external agent. A negative feedback based on the
protein $I\kappa B$ whose expression is controlled by $NF$-$\kappa
B$ is known to be responsible for these oscillations. We study
$NF$-$\kappa B$ oscillations, which have been observed both for
cell populations by Hoffmann et al. (2002)  and for single cells
by Nelson et al. (2004). In order to study cell-to-cell
variability we use Gillespie's algorithm, applied to a simplified
version of the model  proposed by Hoffmann et al. (2002). We
consider the amounts of cellular $NF$-$\kappa B$ and activated
$IKK$ as external parameters. When these are fixed, we show that
intrinsic fluctuations are small in a model with strong transcription,
as is the case of the Hoffmann et al. (2002) model, whether
transcription is quadratic or linear in the number of $NF$-$\kappa B$
molecules. Intrinsic fluctuations can however be large when transcription
is weak, as we illustrate in a model variant. The effect of extrinsic 
fluctuations can be significant: 
cell-to-cell fluctuations of the initial amount of
cellular
$NF$-$\kappa B$ affect mainly the amplitude of nuclear $NF$-$\kappa B$
oscillations, at least when transcription is linear in the
number of $NF$-$\kappa B$ molecules, while fluctuations in the amount of
activated $IKK$
affect both their amplitude and period, whatever the mode of 
transcription. In this case model results are in
qualitative agreement
with the considerable
 cell-to-cell variability
of $NF$-$\kappa B$ oscillations observed by Nelson et al. (2004).

\medskip

\noindent


\end{abstract}

\section{Introduction}

For a number of biological processes cell-to-cell variability
can be very large, even for clonal populations, as pointed
out by Dower and Qwarnstrom (Dower and
Qwarnstrom, 2003), who studied immune response signaling systems
 connected to the family of TIR (Toll/interleukin-1)
receptors. In such cases average cell behavior is a poor
indicator of the way individual cells respond. When for example
cell expression as a function of inducer or enhancer is stochastic,
graded average response can hide all or nothing single cell
response, as shown in a number of studies ~(Fiering et al., 1990,
2000; Hume, 2000; Ko, 1992; Nair et al., 2004). It is therefore
important that in
both experiment and modeling cell-to-cell
variability be investigated. In particular, one needs to be
attentive to this
fact when modeling a biological system
with a set of chemical rate equations for concentrations: these by
definition encapsulate average behavior only. A case in
point
 is provided by recent experiments and modeling of
$NF$-$\kappa B$ activation as a result of $TNF\alpha$ stimulation
of
human Jurkat T cells, U937 monocytes, and mouse fibroblasts
~(Hoffmann et al., 2002), or human cervical carcinoma and
type-S neuroblastoma cells ~(Nelson et al., 2004). Hoffmann et al.
(2002) measure average cell behavior and model it with a set of
chemical rate equations for concentrations, whereas Nelson et al.
(2004) examine the same model but give experimental results
on single cells only,
some of which show large cell-to-cell variations. It is difficult to
reconcile well defined average oscillatory behavior with cell-to-cell
variability
in both amplitude and period of oscillations. Our aim is to understand
(within the confines of the model of Hoffmann et al. (2002))
the origin of the observed ~(Nelson et al., 2004) cell-to-cell
fluctuations of nuclear $NF$-$\kappa B$ oscillations.

 The
fluctuations which underlie cell-to-cell variability can be
separated into intrinsic and extrinsic ~(Elowitz et al., 2002;
Raser and O'Shea, 2004).
Intrinsic fluctuations are
those which result from random occurrence of a defined set of reactions at
random
times, leading to cell-to-cell variability for genetically
identical cells in identical, fixed environments.
Extrinsic fluctuations can have
multiple origins, such as fluctuations in
cytoplasmic and nuclear volumes, in the initial state of the cell,
 signaling cascade fluctuations, or variations in the number of
regulatory molecules.

We should point out that $NF$-$\kappa B$ oscillations are
 a transient phenomenon, unlike the oscillatory process
involved in circadian rhythms.
 Under persistent stimulation of the cell, activation of IKK
itself is temporary, either through adaptation ~(Hoffmann et al.,
2002) or through the action of inhibitory factor A20 ~(Lipniacki
et al., 2004). Thus mathematically the cell which is in a steady
state before stimulation returns to a steady state some time (many
hours) after stimulation. The amplitude and period of oscillations
change as time progresses, and they disappear ultimately.

We use stochastic simulations (Gillespie, 1977) of the rate equations and
show the following:

\noindent
- intrinsic fluctuations are very small for the model parameters used in
the literature
(Hoffmann et al., 2002; Nelson et al., 2004), which correspond to
strong transcription, and they affect only late
nuclear $NF$-$\kappa B$
oscillations. Intrinsic fluctuations can however be large if promoter 
binding is
weak, with rate constants for transcription and translation significantly 
different from those used by Hoffmann et al. (2002)
\\
- extrinsic fluctuations can be significant: fluctuations in the initial
 amount of activated IKK lead
to both amplitude and periodicity fluctuations of $NF$-$\kappa B$
oscillations, which are
features shown by the single cell data of Nelson et al. (2004). This 
result, for the model parameter set of Hoffmann et al. (2202), is 
independent of whether transcription is linear or quadratic in the 
amount of nuclear $NF$-$\kappa B$. However for fluctuations 
in the initial amount of 
cellular $NF$-$\kappa B$
cell-to-cell variability depends on the mode of transcription: 
only for quadratic transcription are both 
amplitude and period affected. For linear transcription only the amplitude 
of oscillations varies. 

These results on cell-to-cell variability make clear the importance of 
determining experimentally the form of the transcriptional relationship 
between 
$NF$-$\kappa B$ and $I\kappa B \alpha$, and the actual values of 
transcriptional 
and translational rate constants.

In the next section we describe the model. This is followed by a section
of
results. Here we
discuss intrinsic noise
and study a variant of the Hoffmann et al. (2002) model with a
different transcription mechanism in order to
gain more insight into the generation of intrinsic noise.
We also describe the results of fluctuations in the initial amount
of $NF$-$\kappa B$, and in the number of activated $IKK$ molecules.
  We conclude with a
discussion in which we speculate on the functional role of a
threshold in the number of $NF$-$\kappa B$ oscillations, as cell
stimulus, and therefore the amount of activated IKK decreases in
intensity.

\section{Description of model}

$NF$-$\kappa B$ (nuclear factor $\kappa B$) plays an important
regulatory role in cell immune response and survival (for a
review, see Tian and Brasier, 2003).
Under normal conditions, it is sequestered in the cell cytoplasm
through binding to $I\kappa B$. When the cell is stimulated
through
for example $TNF\alpha$, the ensuing signaling cascade leads to
activation of IKK, which in turn through binding to $NF$-$\kappa
B$
and $I\kappa B$, leads to ubiquitination of $I\kappa B$, and the
freeing of $NF$-$\kappa B$. $NF$-$\kappa B$ enters the cell
nucleus
(Carlotti et al., 1999),
and regulates the expression of a number of genes, in particular
the gene expressing $I\kappa B$.  $I\kappa B$ protein
enters the
nucleus,
binds to $NF$-$\kappa B$, and that complex is exported. Thus
$NF$-$\kappa B$
which initially was sequestered, then released after IKK
activation, ends up
 by being sequestered again.  As a result, the concentration of
nuclear $NF$-$\kappa B$
shoots up shortly after stimulation, and then decreases because of
the negative feedback associated with $I\kappa B$ expression
through $NF$-$\kappa B$ itself binding to the relevant promoter.
The process can repeat itself if the time scales associated with
nuclear import and export, transcription and translation are
right. Indeed measurements show that the concentration of
nuclear $NF$-$\kappa B$ oscillates over several hours after the
onset
of a lasting stimulation ~(Hoffmann et al., 2002; Nelson et al.,
2004).

For the points we wish to make concerning cell-to-cell
variability,
a pared down version of the model of Hoffmann et al. (2002),
 which only includes the
$I\kappa B$ isoform responsible for $NF$-$\kappa B$ oscillations,
suffices.
The original model includes 64
ordinary differential equations for concentrations, with
reaction rate constants determined or constrained from experiment,
and others constrained by fits to the data obtained with the use
of GEPASI software ( Mendes, 1993).

There are three  isoforms of $I\kappa B$, the main one being
$I\kappa B\alpha$, whose transcription depends strongly on the
presence of nuclear $NF$-$\kappa B$. For the other two,
transcription is not regulated by $NF$-$\kappa B$; Hoffmann et al.
(2002) were interested in elucidating the role of these other two
isoforms in damping $NF$-$\kappa B$ oscillations. Therefore their system
of equations is extensive,
each
isoform participating in the same set of reactions (except for
$NF$-$\kappa B$ dependent transcription), but without any
interaction between isoforms themselves. In our simplified version
(the number of reactions is about 1/3 of those of the original
model), we will only consider the dominant isoform $I\kappa
B\alpha$, which is solely responsible for the occurrence of
$NF$-$\kappa B$ oscillations, and the only one whose knockout is
lethal~(Gerondakis et al., 1999).

Since we are interested in intrinsic noise issues, a model based
on ordinary differential chemical rate equations is inadequate.
We use Gillespie's algorithm (Gillespie, 1977, 2001) to study the
impact of including both
the random occurrence of chemical reactions, and
fluctuations in external parameters on the shape of $NF$-$\kappa B$
oscillations.

We should mention another model of $NF$-$\kappa B$
oscillations ~(Lipniacki et al., 2004), which has as its starting
point the Hoffmann et al. (2002) model, but focuses in addition
on the role of A20, which is regulated by $NF$-$\kappa B$, and
acts
as an inhibitor of IKK, itself activated as a result of cell
stimulation by an external agent. There is also a discussion of
$NF$-$\kappa B$ oscillations by Monk (2003) in the context of a model of
ordinary differential equations with time delays.

The reactions we have used in our model are summarized in Table 1
together
with the relevant rate constants, which
correspond to the choices made by Nelson et al. (2004, supplementary
material)) for
the reactions of Hoffmann et al. (2002). Transcription is linear in the
number of $NF$-$\kappa B$ molecules ~(Lipniacki et al., 2004; Nelson et
al., 2005), rather than quadratic as in the original model ~(Hoffmann et
al., 2002).  This change, from quadratic to linear,
diminishes the sensitivity of $NF$-$\kappa B$ oscillations to
fluctuations in the initial amount of $NF$-$\kappa B$ present in a
cell (Barken et al., 2005; Nelson et al., 2005). We will give results
for both choices of transcription in the next section. The model as it
stands  still has some
unsatisfactory features, namely
the amount of transcribed molecules rises linearly (or
quadratically as in Hoffmann et al. (2002))
with the number of $NF$-$\kappa B$ molecules entering the cell nucleus,
without any
 mechanism of saturation, and the number of
$I\kappa B\alpha$ mRNA molecules is excessive, of the order of a thousand
in the
steady state (for the simulation underlying Figure 1), rising to
much higher numbers (several times ten thousand) once IKK is
activated. We have however checked, as we will discuss in section
3.1.,
that as a result of decreased
transcription and a corresponding  increased translation, the
number of $I\kappa B\alpha$ mRNA molecules is reduced, while
$NF$-$\kappa B$ oscillations themselves are not
affected, because the amount of $NF$-$\kappa B$ present in the
cell is large; the intrinsic noise in the system, however, becomes larger,
the more so as the transcription rate decreases.

In the model, $NF$-$\kappa B$ is not degraded, and therefore the
number of $NF$-$\kappa B$ molecules remains constant through the
formation and disintegration of complexes in which it
participates. We use cells with volumes of 1000 $\mu^3$, divided
equally between cytoplasm and nucleus. We take an amount of
 $0.05\mu M$ relative to the total cell volume for $NF$-$\kappa B$.
Thus the number of $NF$-$\kappa B$ molecules initially
introduced, corresponding to $0.05 \mu M$ for the whole cell, is
30,000. After
an equilibration period of
3000 minutes, we follow Hoffmann et al. (2002) in introducing an
equal amount of IKK (0.05 $\mu M$), which
represents activation of the signaling pathway through cell
exposure to $TNF\alpha$.

The shape of $NF$-$\kappa B$ oscillations will for a given
cell volume depend on (all other parameters being fixed) the
fraction of volume occupied by the nucleus ~(Lipniacki et al., 2004),
because reactions take
place in both cytoplasm and nucleus and there is nuclear import
and export. Their shape depends as well on the comparative amount of
initial
$NF$-$\kappa B$ and IKK activated after cell stimulation.

$NF$-$\kappa B$ oscillations, which are
more or less damped, are sensitive to a number of model parameters
such as the rate of transcription of $I\kappa B\alpha$ or its
translation, or $I\kappa B\alpha$ mRNA degradation, which are unknown
and determined from fits to experiments. Thus it is not always
possible to evaluate the significance of the precise form of the
oscillations from comparing model results with experiment.

\section{ Results.}

Data are obtained in different cell lines (Hoffmann et al., 2002;
Nelson et al., 2004) with different techniques and therefore
comparisons can be hazardous. Nevertheless, the data on
$NF$-$\kappa B$ signaling show well defined oscillations ( of
period close to 100 minutes) when averaged over cells, when only
isoform $I\kappa B\alpha$ is present (Hoffmann et al., 2002,
supplementary material, fig. S1). However the single cell data of
Nelson et al. (2004) show large cell-to-cell variations in both
amplitude and periodicity, which if typical, would be expected to
broaden or even wash out average $NF$-$\kappa B$ oscillations. It
is in this context that we will next discuss our results. We  will
examine three possible sources of cell-to-cell variability:
intrinsic fluctuations on one hand, and on the other extrinsic
fluctuations  in either the initial number of $NF$-$\kappa B$
molecules or of activated $IKK$ molecules.

\subsection{Intrinsic noise.}

Intrinsic noise is small in the models of Hoffmann et al. (2002)
and Nelson et al. (2004) given the rate constants they
employ. Whether transcription is linear or quadratic in the number
of $NF$-$\kappa B$ molecules, intrinsic fluctuations become
significant only about 600 minutes after $IKK$ activation. In
Figure 1 we show, for the case of linear transcription, and for
the parameters of Table 1, nuclear $NF$-$\kappa B$ oscillations as
a function of time after activation by $IKK$ after 3000 minutes of
equilibration. For this figure the initial numbers of $NF$-$\kappa
B$ and activated $IKK$ molecules are fixed. Thus any fluctuation
is purely intrinsic, due to the random order in which reactions
occur in any given cell. The first three or four oscillations have
a period of about 100 minutes, thereafter the period lengthens.
The figure shows average behavior and the results for three
randomly chosen single cells. The comparison illustrates the point
that fluctuations are very small: in each cell oscillations for 8
hours after activation are almost identical to the average. (The
rapid small amplitude fluctuations over times less or much less
than a minute are washed out in the representation of Figure 1.)
Thus cell-to-cell variability is very small. The main reason is
that the number of molecules is large: with values of the order of
0.05 $\mu M$ for initial numbers of $NF$-$\kappa B$ and activated
$IKK$ molecules, and cells of volume 1000 ${\mu}^3$, their numbers
are in the tens of thousands. Moreover transcription, the strength
of which increases with the number of $NF$-$\kappa B$ molecules,
is strong. If in the models considered, one decreases, while
keeping all other rate constants unchanged, the rate of
transcription by merely a factor of 2 (for linear transcription)
or 4 (for quadratic transcription) , $NF$-$\kappa B$ oscillations
beyond the first one are very weak and disappear.

Nevertheless the strength of intrinsic fluctuations indeed varies with the
values of the rate constants. As already mentioned in Section 2, we have
decreased, within the model with linear transcription, the  rate
of transcription  of $I\kappa B\alpha$ mRNA by a factor of 10, in order
to have fewer mRNA, and increased translation by the same factor,
thus keeping the number of $I\kappa B\alpha$ proteins essentially
unchanged.
 This model still yields good oscillations in nuclear
$NF$-$\kappa B$ but with increased cell-to-cell variability, as shown in
Figure 2, and as compared to the single cell curves in Figure 1.
The decreased copy number of $I\kappa B\alpha$ mRNA thus
leads to noise
in its own oscillations that is reflected in the behavior of
$NF$-$\kappa B$. The
intrinsic noise is increased (up to factors of two) over the
original model.  We will investigate further the connection between
strength of transcription and intrinsic noise
 in the following section with a more appropriate model which includes
explicit promoter binding.

\subsubsection{Model variant with DNA.}

The model of Hoffmann et al. (2002)
treats transcription in an oversimplified way without including
DNA. In order to examine the effect of the stochastic binding and
unbinding of  $NF$-$\kappa B$
 to the promoter site of the DNA on the level of intrinsic noise
 we have considered the following model: the transcription of the
 mRNA coding for $I\kappa B\alpha$, denoted by \emph{Ictr}, occurs
 through the reactions.
 \begin{eqnarray}
DNA\,+\,Nn&\eqwithrates{kf}{kb}&DNA^*\\
DNA^*&\stackrel{tr'_2}{\longrightarrow}&DNA\,+\,Nn\,+\,Ictr
\end{eqnarray}
Since the sum of the concentrations of the $DNA$ and the complex
$DNA^*$ is conserved, we have, $c_D\,+\,c_{D^*}\,=\,c_0$ where
$c_0\,=\,2/300000\, \mu M$ given the volume of the nucleus ($500
\mu^3$) and assuming a diploid cell.  If the first reaction above
occurs rapidly and is hence, in equilibrium, we can combine the
two equations in the spirit of Michaelis-Menten to obtain
$$Nn\,\stackrel{tr_{2,eff}}{\longrightarrow}\,Nn\,+\,Ictr$$
where the effective transcription rate that depends on the nuclear
NF$\kappa B$ concentration $Nn$ is given by
$$tr_{2,eff}\,=\,\frac{c_0\,tr'_2}{\frac{kb}{kf}\,+\,Nn}\,.$$
Since the maximum value of $Nn$ is $0.1 \mu M$ we chose
$kb/kf\,=\,0.2\,\mu M$; changing the value up to $1.0\,\mu M$ had
negligible effect on the results given below. In order to obtain
 similar transcription rates as for the linear model we chose
$tr'_2\times c_0$ to be equal to the value of $tr_2$ used in the
original model given in Table 1. In order to ensure that the
binding and unbinding of $Nn$ occur sufficiently rapidly we chose
$kb\,=\,1200\,min^{-1}$ and the translation rate $tr1$ is
increased by a factor of $2$ from the value given in Table 1.
With this high rate of promoter binding, one is in a
situation, as for the models of linear or quadratic transcription
~(Hoffmann et al., 2002; Nelson et al., 2004), where the  relevant
promoter is strongly activated and transcriptional stochasticity is small
(Raser and O'Shea, 2004). This is confirmed by the simulation
reported in Figure 3, which shows the concentration
and standard deviation of nuclear $NF$-$\kappa B$ as a function of
time after activation by $IKK$ after 3000 minutes of
equilibration. The noise is
small in the first four
oscillations and increases at larger times, in the same way as shown in
Figure 1 for linear transcription with the parameter set of Table 1.

{\bf Transcriptional noise.}

Moving away from the regime of strong transcription embodied in
the models with linear or quadratic transcription ~(Hoffmann et
al., 2002; Nelson et al., 2004), one can now explore what happens
when promoter binding gets weaker, with a rate that gets closer to
the rates which determine $I\kappa B\alpha$ protein
ubiquitination, namely rates $r1$ and $r4$ (cf. Table 1).
Transcriptional fluctuations are then no longer averaged out over
the corresponding time scales ~(Lewis, 2003). One can do so since,
contrary to most rate constants which are determined or
constrained by experiment ~(Hoffmann et al., supp. material,
2002), those of transcription and translation, $tr2$ and $tr1$
respectively (cf. Table 1), result from fits to the oscillation
data. However, as one decreases the rate of promoter binding, it
is necessary to increase correspondingly the rate of translation
in order to maintain well defined nuclear $NF$-$\kappa B$
oscillations. Figure 4 illustrates  the impact of weak
transcription on the strength of intrinsic noise, for a binding
rate of $kb=12 min^{-1}$, a hundredfold decrease as compared to
the case of strong transcription (see above), and a corresponding
200 fold increase of the translation rate $tr1$ of Table 1. The
intrinsic noise is much more significant than in Figure 3, and, as
shown in Figure 5, it affects both the amplitude and periodicity
of $NF$-$\kappa B$ oscillations. Indeed, the intrinsic noise is
sufficiently large that
 the oscillations obtained from averaging the stochastic simulations is
more damped than those obtained from the deterministic system.
This underscores the potential pitfalls of deducing rate constants
by fitting with deterministic models.

 One could investigate the effects of further
reducing promoter binding. For our purpose which is to establish
and clarify the possible sources of the observed cell-to-cell
variability (Nelson et al., 2004) of $NF$-$\kappa B$ oscillations,
it is sufficient to emphasize that intrinsic noise can be
significant, as important as extrinsic noise, if we do not confine
ourselves to the fitted values of transcription and translation
rates of the set of reactions considered by Hoffmann et al.
(2002).


\subsection {Extrinsic noise.}

Since the data indicate large cell-to-cell fluctuations
(Nelson et al., 2004), these must arise from fluctuations in
external parameters, at least for the models with strong transcription,
for which, as shown above, intrinsic noise is small.
 We consider two possible sources here, the number of initial
$NF$-$\kappa B$ present in a given cell, which varies between
cells in an
analogous situation (Dower and
Qwarnstrom, 2003),
and the amount of $IKK$, which can vary due to signaling cascade
fluctuations.
 The models considered are those with transcription quadratic in the
number of
$NF$-$\kappa B$ molecules ~(Hoffmann et al., 2002) and its version with
linear transcription ~(Nelson et al., 2004) with the rate constants 
of Table 1. For quadratic
transcription, the linear transcription reaction in Table 1 is replaced
by $2
Nn \rightarrow 2 Nn + Ictr$ with rate constant $tr2= 1.027125 \mu M^{-1}
min^{-1}$. For Figure 7 moreover rate constants $tr1$ and $tr3$ of Table
1 are multiplied by 2.

\subsubsection {$NF$-$\kappa B$ fluctuations.}

As alluded to in section 2, there is a significant difference here
between the models with linear and quadratic transcription
~(Barken et al., 2005; Nelson et al., 2005). For linear
transcription, moderate fluctuations in the amount of $NF$-$\kappa
B$(see below) affect the amplitude only of single cell
oscillations, thus maintaining well-defined average oscillations.
For quadratic transcription, both amplitude and period of single
cells fluctuate, leading to reduced or washed out average
oscillations. The following discussion explains this point.

In Figure 6, for linear transcription,  are shown results for
$NF$-$\kappa B$ oscillations for the average over many cells for a
Gaussian distribution of initial number of $NF$-$\kappa B$ (with
average 30000, $\sigma=$5000) (black curve) and single cell
behavior when initial $NF$-$\kappa B$ is one or two $\sigma$ away
from the average. IKK remains fixed in all cases at IKK=30000.
Average $NF$-$\kappa B$ oscillations in Figure 6 are of course
very similar to those in Figure 1, diverging from them only beyond
the fourth oscillation, when amplitudes become smaller. As
expected, single cell behavior differs from the average only in
amplitude, significantly so for $2\sigma$ deviations from average.
But periodicity is maintained for 8 hours after $IKK$ activation,
and thus, insofar as the data of Nelson et al. (2004) show
significant early variation in oscillation period from cell to
cell, there have to be other sources of variability in a model
with strong {\it linear} transcription.

On the contrary, for quadratic transcription, fluctuations in the
initial number of $NF$-$\kappa B$  molecules, affect both
amplitude and periodicity of oscillations, not just their
amplitude as in the case of linear transcription. The result is
that contrary to what happens in Figure 6 for linear
transcription, where oscillations remain clear and distinct, for
an initial Gaussian distribution of $NF$-$\kappa B$ over cells (at
a fixed number of activated IKK), here oscillations in the number
of nuclear $NF$-$\kappa B$ oscillations, are very much damped out
as shown in Figure 7, as a result of cell-to-cell variability
in both amplitude and period of oscillations.

\subsubsection {$IKK$ fluctuations.}

Within the model parameters considered, we turn to studying
fluctuations in the initial number of IKK which activates the
cellular oscillatory response. Whatever the type of transcription,
linear or quadratic, fluctuations in the initial amount of
activated $IKK$ lead to strong cell-to-cell variability. We show
here our results for the case of linear transcription.

 Figure 8 shows, for  fixed $NF$-$\kappa B$ = 30000,
the average over many cells for a Gaussian distribution of the
initial number of IKK (average=30000, $\sigma=5000$), and single
cell response for IKK=25000 and IKK=35000. Several points can be
made here:

- when the initial  number of IKK activated by $TNF\alpha$ fluctuates, not
only does the amplitude of $NF$-$\kappa B$ oscillations vary, but
also their periodicity

- as a result, single cell fluctuations are washed out in the
average. Average response over the cell population represents poorly
single cell response beyond the first
oscillation

- the difference between maxima and minima of the oscillations
gets larger as IKK decreases from its average, while at the same
time oscillatory period increases. Possible implications of the
latter result are given in the discussion.

It is clear from the figures that whatever the source of cellular
fluctuations, cell-to-cell variability typically increases with
time after the onset of IKK activation. This can be illustrated
further by computing the standard deviation of the noise in the
number of nuclear NF-$\kappa$B. In Figure 9 we compare standard
deviations as a function of time for the case of Figure 8, when
IKK fluctuates according to a Gaussian of average and width of
respectively equal 30000 and 5000 molecules, at fixed initial
concentration of $NF$-$\kappa B$ of 30000 molecules (0.05 $\mu
M$).  On one hand there is
$\sigma_{tot}^2=\overline{<Nn^2>}-{\overline{<Nn>}}^2$, the total
variance of nuclear $NF$-$\kappa B$ oscillations, on the other
hand there is $\sigma_{int}^2=\overline{<Nn^2>-<Nn>^2}$, the
intrinsic variance.
 Here $Nn$ denotes nuclear $NF$-$\kappa B$, the brackets denote
average over intrinsic fluctuations, and the overline average over
fluctuations in the external parameter (Elowitz et al., 2002;
Swain et al., 2002). We observe (Fig. 9) that $\sigma_{tot}$,
which includes extrinsic fluctuations due to fluctuations in IKK,
is much larger than $\sigma_{int}$ except for large times ( 13
hours) after IKK activation. Thus for most of the time after IKK
activation, extrinsic fluctuations dominate. The magnitude of
$\sigma_{tot}$ shortly after IKK activation reflects the
significance of IKK fluctuations in determining cell-to-cell
variability in terms  of both amplitude and period of $NF$-$\kappa
B$ oscillations.

 In this subsection, we have studied the effect of varying 
the initial
amount of activated $IKK$ around an average value, which we denote by 
$IKK_0$,  while keeping the 
number
of $NF$-$\kappa B$ molecules fixed. Let us now comment on how average 
oscillations change when $IKK_0$ changes significantly compared to the 
value considered up to now. When $IKK_0$ decreases 
by $50\%$ compared to the value chosen by Hoffmann et al. (2002),
oscillations for both linear and quadratic transcription remain
strong with an average increase in period of 20 to 30 percent; for
an increase of $50\%$ in $IKK_0$, the period of oscillations
decreases for quadratic transcription, whereas for linear
transcription oscillations are significantly damped beyond the
first one. In the experiments discussed (Hoffmann et 
al., 2002; Nelson et
al., 2004) there are no results for the effect of varying the
amount of the external stimulus, $TNF\alpha$, on  $NF$-$\kappa B$
oscillations. We note that the utility of our predictions for 
the average oscillations as a function of activated $IKK$
depends on knowing  how the
amount of activated $IKK$ is related to the amount of external
stimulant.

\section{ Discussion.}

The data of Hoffmann et al. (2002) and Nelson et al. (2004) on
$NF$-$\kappa B$ oscillations are seemingly contradictory. Hoffmann et al.
(2002,
suppl. material, fig. S1) observe well-defined oscillations for a
population of cells, whereas Nelson et al. (2004) observe (for the
identical stimulus) large cell-to-cell variability, in both amplitude and
period of oscillations, which would normally lead to washed out average
oscillations. However both groups use different cell lines, and
experimental methods differ (for a discussion of these see the technical
comment of Barken et al. (2005) and the response by Nelson et al. (2005)).

From a modeling point of view, for the class of reactions considered here
(cf. Table 1), we summarize our key results:\\
- {\it intrinsic} fluctuations are small when transcription is strong
(which is the case of the model of Hoffmann et al.(2002)), whether
transcription is linear or quadratic in the number of $NF$-$\kappa
B$ molecules. However intrinsic fluctuations can be strong, as significant
as any extrinsic noise, if the rate of promoter binding is small, and rate 
constants for transcription and translation of $I\kappa B$ become very 
different from the ones used by Hoffmann et al. (2002); if this is the 
case 
intrinsic noise can account for the observed cell-to-cell variability.
Hence, it is important to determine transcription rate constants 
experimentally. 

In all cases considered, {\it extrinsic} fluctuations can be significant:

- in the model of Hoffmann et al. (2002), fluctuations in the 
amount of activated $IKK$ molecules can account for
the observed cell-to-cell variability (Nelson et al., 2004), because they 
affect both amplitude
and period of $NF$-$\kappa B$ oscillations, whatever the mode of
transcription, linear or quadratic in the number of
$NF$-$\kappa B$ molecules, of $I\kappa B\alpha$.

- as to fluctuations in the amount of $NF$-$\kappa B$ 
molecules, their impact depends on the type of transcription, 
linear or quadratic.
For linear transcription 
the amplitude only of oscillations varies, not their periodicity, whereas
for quadratic transcription, both amplitude and periodicity vary.
 Only in the latter case would these
fluctuations be a possible source of the observed cell-to-cell variability
(Nelson et al., 2004)

Fluctuations in individual cell behavior thus provide
a source of  damping for $NF$-$\kappa B$ population oscillations that is
different from the
one considered by Hoffmann et al. (2002), which depends on the existence
of $I\kappa B$ isoforms besides $I\kappa B\alpha$.

Our results on cell-to-cell variability of $NF$-$\kappa B$ oscillations
(in the context of the model of Hoffmann et al., 2002) help clarify
the experimental issues that need to be resolved for determining its 
origins.
These requirements can be formulated as questions:    
 what is the distribution over the cell
population of the amount of $NF$-$\kappa B$, and of activated
$IKK$; what are the rates of transcription and translation; is
transcription linear or quadratic (in the right range) in the
number of $NF$-$\kappa B$ molecules? When answers to these questions
will be forthcoming, one will be able to determine, within a given model,
the sources of cell-to-cell variability of $NF$-$\kappa B$ oscillations,
and eventually reconcile 
 the well-defined average
oscillations observed by Hoffmann et al. (2002) and the large cell-to-cell
variability of Nelson et al. (2004) as a matter of differences in the 
amount of fluctuations due to cell type, cell 
environment or experimental method.



The key biological issue is the functional role of $NF$-$\kappa B$ 
oscillations, which is not known.
Since $NF$-$\kappa B$ participates in regulating the expression
of many genes, its oscillations can be  transferred to the
corresponding proteins. One can speculate that
the number of $NF$-$\kappa B$
oscillations in a given time period could play the role of a
threshold for setting into motion many aspects of cellular
response. Consider Figure
8: as the level of $IKK$ goes down the period of oscillations
lengthens. Assume that four oscillations in the first 400
minutes after IKK activation represent the threshold below
which signaling based on $NF$-$\kappa B$ oscillations becomes
ineffective. In Figure 8 , the case $IKK$=25000 is marginal
 ($NF$-$\kappa B$ is fixed). Thus as the amount of $IKK$ goes lower,
say to $IKK$ =20000, and if this amount represents the average
amount of $IKK$ activated by an outside agent, only
fluctuations in $IKK$  will enable some cells to reach the
threshold of four oscillations in the required time, and thus
respond to the external stimulation. This effect of stochasticity
would lead to cellular response even for small, below threshold,
doses of
external agent, as might happen for immune response to a
pathogen.

\vspace{0.5cm}

{\bf Acknowledgment.} We thank Stuart Sealfon for
interesting discussions.

\newpage

\section {References}

\medskip
\noindent Barken, D., Wang, C.J., Kearns, J., Cheong, R.,
Hoffmann, A., and Levchenko, A., 2005. Comment on "Oscillations
 in $NF$-$\kappa B$ signaling control the dynamics of gene
expression". Science 308, 52a.

\medskip
\noindent
Carlotti, F., Chapman, R., Dower, S.K., and Qwarnstrom, E.E.,
1999. Activation of nuclear factor $\kappa B$ in single living
cells. J. Biol. Chem. 274, 37941-37949.

\medskip
\noindent Dower, S.K., and Qwarnstrom, E.E., 2003. Signalling
networks, inflammation and innate immunity. Bioch. Soc. Trans. 31,
1463-1467.

\medskip
\noindent Elowitz, M. B., Levine, A.J., Siggia, E.D.,
vvand
Swain, P.S., 2002. Stochastic gene expression in a single cell.
Science 297, 1883-1886.

\medskip
\noindent  Fiering, S., Northrop, J.P., Nolan, G.P.,
Mattila, P.S.,
Crabtree, G.R., and Herzenberg, L.A., 1990.
Single cell assay of a transcription factor reveals a threshold in
transcription activated by signals emanating from the T-cell
antigen
receptor. Genes Dev. 4, 1823-1834.

\medskip
\noindent  Fiering, S., Whitelaw, E., and Martin, D.I.K.,
2000. To
be or not to be active:the stochastic nature of enhancer action.
BioEssays 22, 381-387.

\medskip
\noindent Gerondakis, S., Grossmann, M., Nakamura, Y., Pohl, T.,
and Grumont, R., 1999. Genetic approaches in mice to understand
Rel/$NF-\kappa B$ and $I\kappa B$ function:transgenics and
knockouts. Oncogene 18, 6888-6895.

\medskip
\noindent Gillespie, D.T., 1977. Stochastic simulations of coupled
chemical reactions. J. Phys. Chem. 81, 2340-2361.

\medskip
\noindent Gillespie, D.T., 2001. Approximate accelerated
stochastic stimulation of chemically reacting systems.
 J. Chem. Phys. 115, 1716-1733.

\medskip
\noindent Hoffmann, A., Levchenko, A., Scott, M.L., and Baltimore
D., 2002. The $I\kappa B-NF-\kappa B$ signaling module: temporal
control and selective gene activation. Science 298, 1241-1245.

\medskip
\noindent  Hume, D. A. 2000. Probability in transcriptional
regulation and its implication for leukocyte differentiation and
inducible gene expression. Blood 96, 2323-2328.

\medskip
\noindent  Ko, M.S. 1992. Induction mechanism of a single gene
molecule: Stochastic or Deterministic?
BioEssays 14, 341-346.

\medskip
\noindent Lewis, J. 2003. Autoinhibition with transcriptional delay: a
simple mechanism for the zebrafish somitogenesis oscillator. Current
Biology 13, 1398-1408.

\medskip
\noindent Lipniacki, T., Paszek, P., Brasier, A.R., Luxon, B., and
Kimmel, M., 2004. Mathematical model of $NF-\kappa B$ regulatory
module. J. Theor. Biol. 228, 195-215.

\medskip
\noindent Mendes, P., 1993. GEPASI: a software package
for modelling the dynamics, steady state and control of
biochemical and other systems. Comp. Appl. Biosci. 9, 563-571.

\medskip
\noindent Monk, N.A.M. 2003. Oscillatory expression of Hes1, p53,
and $NF$-$\kappa B$ driven by transcriptional time delays.
Current Biology 13, 1409-1413.

\medskip
\noindent Nair, V.D., Yuen, T., Olanow, C.W., and Sealfon, S.C.,
2004. Early single cell bifurcation of pro- and antiapoptopic
states during oxidative stress. J. Biol. Chem. 279, 27494-27501.

\medskip
\noindent Nelson, D.E., Ihekwaba, A.E.C., Elliott, M., Johnson,
J.R., Gibney, C.A., Foreman, B.E., Nelson, G., See, V., Horton,
C.A., Spiller, D.G., Edwards, S.W., McDowell, H.P., Unitt, J.F.,
Sullivan, E., Grimley, R., Benson, N., Broomhead, D., Kell, D.B.,
and White M.R.H., 2004. Oscillations in $NF-\kappa B$
signaling control the dynamics of gene expression. Science 306,
704-708.

\medskip
\noindent Nelson, D.E., Horton, C.A., See, V., Johnson,
J.R., Nelson, G., Spiller, D.G., Kell, D.B.,
 and White M.R.H., 2005.
Response to Comment on "Oscillations
 in $NF-\kappa B$ signaling control the dynamics of gene
expression". Science 308, 52b.

\medskip
\noindent Raser, J.M., and O'Shea, E.K., 2004. Control of
stochasticity in eukaryotic gene expression. Science 304,
1811-1814.

\medskip
\noindent Swain, P.S., Elowitz, M.B., and Siggia, E.D., 2002.
Intrinsic and extrinsic contributions to stochasticity in gene
expression. Proc. Nat. Ac. Sci. 99, 12795-12800.

\medskip
\noindent Tian, B., and Brasier, A.R., 2003. Identification of a
nuclear factor kappa B-dependent gene network. Recent Prog. Horm.
Res. 58, 95-130.

\pagebreak

\begin{table}[H]
\caption{Reactions and Rate constants}
\[
\begin{tabular}{||c|c|c||} \hline
Rate\, constant & Reaction & Value
\\
\hline $a_4 $  & $Ic+Nc \rightarrow INc $ & $30 \mu M^{-1}min^{-1} $ \\
\hline $d_4 $  & $INc \rightarrow Ic+Nc$ & $0.03 min^{-1} $ \\
\hline $a_7 $  & $IKK+INc\rightarrow INcp$   & $ 11.1 \mu M^{-1}min^{-1}$
\\
\hline $d_1 $  & $ INcp \rightarrow IKK+INc$   & $0.075 min^{-1} $ \\
\hline $r_4 $  & $INcp \rightarrow IKK+Nc $  & $ 1.221 min^{-1}$ \\
\hline $a_1 $  & $IKK+Ic\rightarrow Icpr   $  & $1.35 \mu
M^{-1}min^{-1}$\\
\hline $d_1 $  & $Icpr\rightarrow IKK+Ic $   & $ 0.075 min^{-1} $ \\
\hline $a_4 $  & $ Icpr +Nc \rightarrow INcp$  & $ 30 \mu M^{-1}min^{-1}$ \\
\hline $d_4 $  & $INcp\rightarrow Icpr +Nc $  & $0.03 min^{-1} $ \\

\hline $r_1 $  & $Icpr\rightarrow IKK  $  & $ 0.2442 min^{-1} $ \\
\hline $deg_1 $  & $Ic\rightarrow \emptyset  $  & $ 0.00675 min^{-1} $ \\
\hline $tp_1 $  & $ Ic \rightarrow In $  & $ 0.018 min^{-1} $ \\
\hline $tp_2 $  & $ In \rightarrow Ic $  & $ 0.012 min^{-1} $ \\
\hline $k_1 $  & $ Nc \rightarrow Nn $  & $ 5.4 min^{-1} $ \\
\hline $k_{01} $  & $ Nn \rightarrow Nc $  & $ 0.0048 min^{-1} $ \\
\hline $k_2 $  & $INn\rightarrow INc  $  & $ 0.8294 min^{-1} $ \\
\hline $tr_2 $  & $Nn\rightarrow Nn+Ictr  $  & $ 0.0582 min^{-1} $ \\
\hline $tr_3 $  & $ Ictr\rightarrow \emptyset $  & $ 0.0168 min^{-1} $ \\
\hline $tr_1$  & $Ictr\rightarrow Ictr+Ic  $  & $ 0.2448 min^{-1} $ \\
\hline $a_4 $  & $In+Nn\rightarrow INn  $  & $ 30 \mu M^{-1}min^{-1} $ \\
\hline $d_4 $  & $INn\rightarrow In+Nn  $  & $ 0.03 min^{-1} $ \\
\hline $k_{02} $  & $ IKK\rightarrow \emptyset$  & $ 0.0072 min^{-1} $ \\
\hline
\end{tabular}
\]
\label{Tab}
Rate constants are from Nelson et al. (2004), based on the model of
Hoffmann et al. (2002).\\

Notation:\\
 Nn= nuclear $NF$-$\kappa B$, Nc= cytoplasmic $NF$-$\kappa B$\\
In=  nuclear $I\kappa B\alpha$, Ic=cytoplasmic $I\kappa B\alpha$\\
INc= cytoplasmic $NF$-$\kappa B$-$I\kappa B\alpha$ complex\\
INn= nuclear $NF$-$\kappa B$-$I\kappa B\alpha$ complex\\
INcp= complex of IKK and INc\\
Icpr= complex of IKK and $I\kappa B\alpha$\\
Ictr= $I\kappa B\alpha$ mRNA

\end{table}

\pagebreak

\clearpage

\section{ Figure captions}

Figure 1. Nuclear $NF$-$\kappa B$ as a function of time for the model with
linear transcription.
IKK activation sets in after 3000 minutes. Initial total amount of
$NF$-$\kappa B$ and amount of IKK are fixed at 0.05 $\mu M$.
The full curve is average behavior, the broken curves show three
randomly chosen single cell oscillations.

Figure 2. Nuclear $NF$-$\kappa B$ as a function of time for the
model with linear transcription. Here the values of rate constants
tr2 for transcription and tr1 for translation are 1/10 and 10
times respectively  the values given in the Table, which are used
for Figure 1 (see text). IKK activation sets in after 3000
minutes. Initial total amount of $NF$-$\kappa B$ and amount of IKK
are fixed at 0.05 $\mu M$. The full and two broken curves
correspond to three randomly chosen single cell
oscillations and show the increased cell-to-cell variability
as compared to Figure 1.

Figure 3. Nuclear $NF$-$\kappa B$ as a function of time for the
model variant described in section 3.1.1. Here $kb=1200 min^{-1}$.
$IKK$
activation sets in
after 3000 minutes. Initial total amount of $NF$-$\kappa B$ and
amount of $IKK$ are fixed at 0.05 $\mu M$. The full curve is the
average behavior of $NF$-$\kappa B$, the broken curve represents
the standard deviation due to intrinsic noise.

Figure 4. Nuclear $NF$-$\kappa B$ as a function of time for the
model variant described  in section 3.1.1. Here $kb=12 min^{-1}$;
transcription is weak.
The full curve is  average cell behavior, the broken
line is the standard deviation $\sigma$.

Figure 5. Nuclear $NF$-$\kappa B$ as a function of time for the
model variant described  in section 3.1.1. Here $kb=12 min^{-1}$;
transcription is weak.
The curves correspond to the behavior of three single cells.
Oscillations differ in both amplitude and phase.

Figure 6. Nuclear $NF$-$\kappa B$ as a function of time for the model
with
linear transcription.
IKK activation sets in after 3000 minutes.
The amount of initially activated IKK is 0.05 $\mu M$.
The full curve represents an average over 1000 single cells
for an initial amount of $NF$-$\kappa B$ following a gaussian
distribution of average number equal to 30000 (corresponding to
0.05 $\mu M$ for the cell volume chosen), and width $\sigma=5000$.
The broken curves correspond to one single cell behavior for,
by decreasing amplitude, $NF$-$\kappa B$= 35000, 25000, and 20000
molecules.

Figure 7. Nuclear $NF$-$\kappa B$ as a function of time for the
model with quadratic transcription. $IKK$ activation sets in after
3000 minutes. The amount of initial activated IKK is 0.05 $\mu M$.
The full curve represents an average over 1000 single cells for an
initial amount of $NF$-$\kappa B$ following a Gaussian
distribution of average number equal to 30000 (corresponding to
0.05 $\mu M$ for the cell volume chosen), and width $\sigma=5000$.

Figure 8. Nuclear $NF$-$\kappa B$ as a function of time for the
model with linear transcription. IKK activation sets in after 3000
minutes. The amount of initial $NF$-$\kappa B$ is fixed at 0.05
$\mu M$. The full curve represents an average over 1000 single
cells for an initial amount of $IKK$ following a Gaussian
distribution of average number equal to 30000 (corresponding to
0.05 $\mu M$ for the cell volume chosen), and width $\sigma=5000$.
The broken curves correspond to single cell behavior for, by
decreasing amplitude, IKK= 25000 and 35000 molecules.

Figure 9. Standard deviations (in $\mu M$) of nuclear $NF$-$\kappa B$
oscillations for the case
considered in Figure 3, as a function of time. IKK activation sets in
 after 3000 minutes. $\sigma_{tot}$ (broken
line)
represents the total noise, $\sigma_{int}$ (full line) the intrinsic
noise (see text).

\pagebreak

\mbox{}

\begin{figure}[htbp]
\centering
\includegraphics{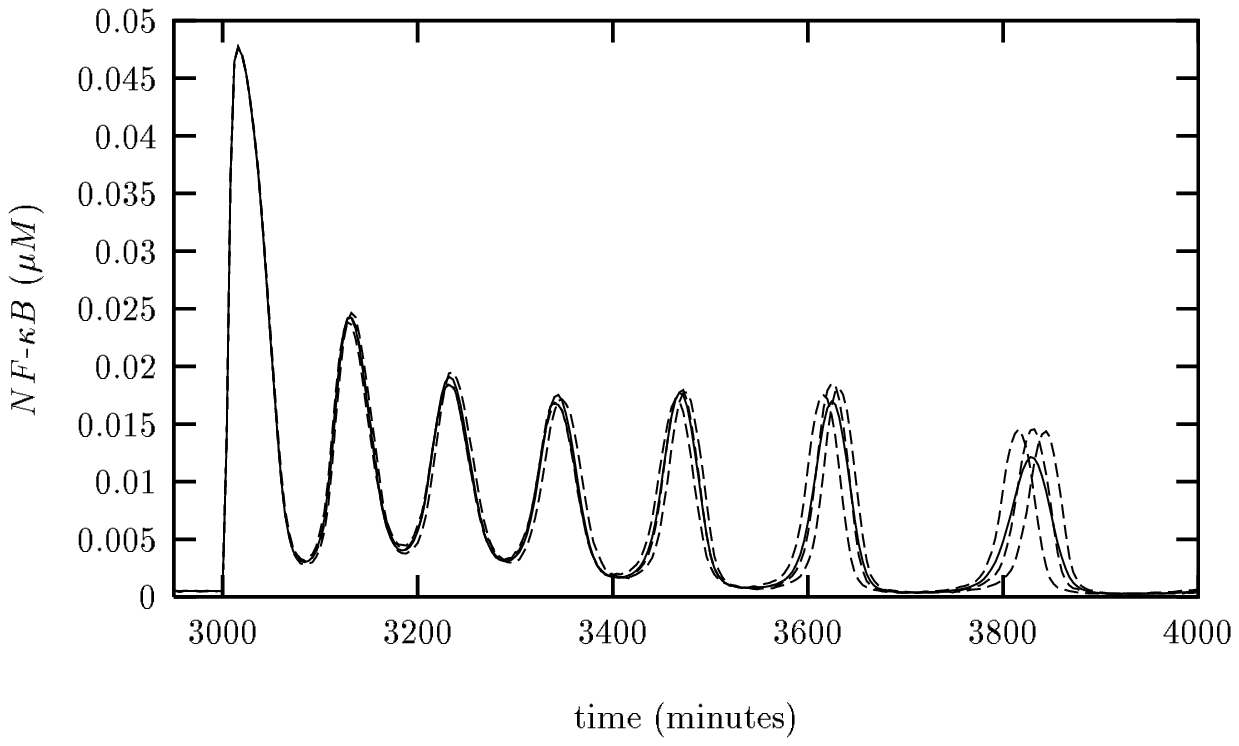}
\caption{}
\end{figure}

\begin{figure}[htbp]
\centering
\includegraphics{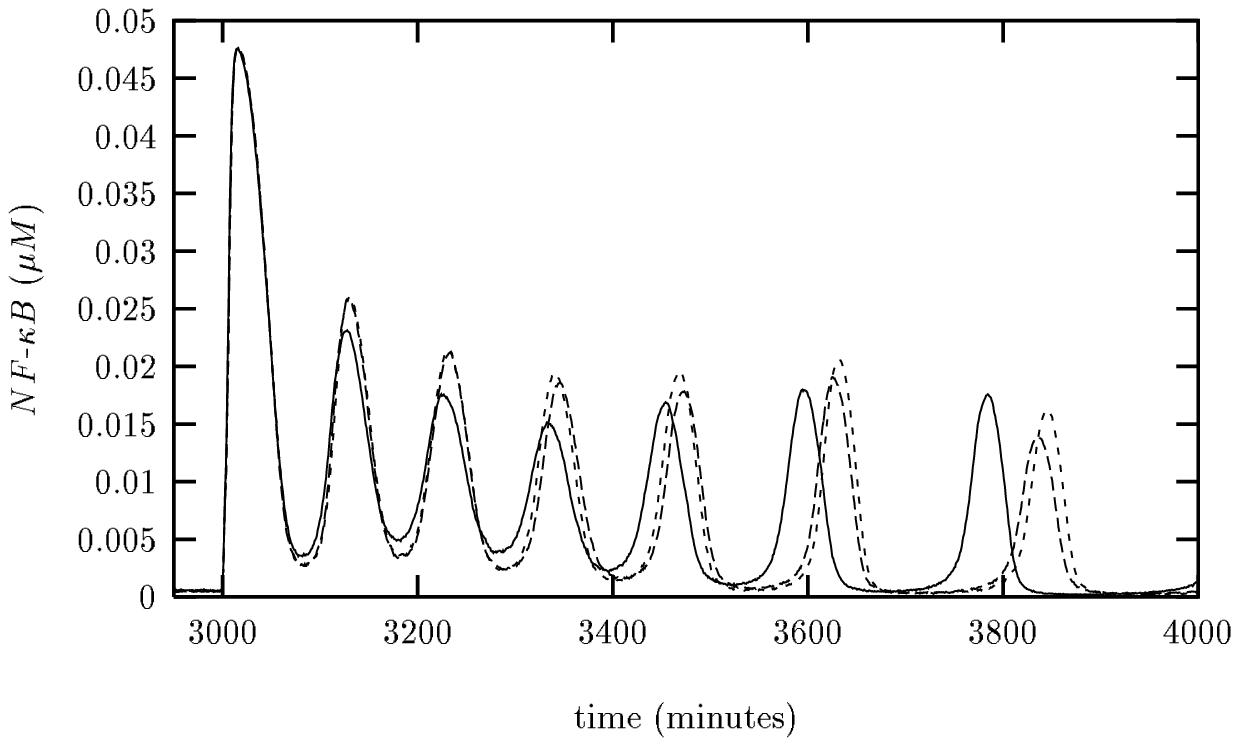}
\caption{}
\end{figure}

\begin{figure}[htbp]
\centering
\includegraphics{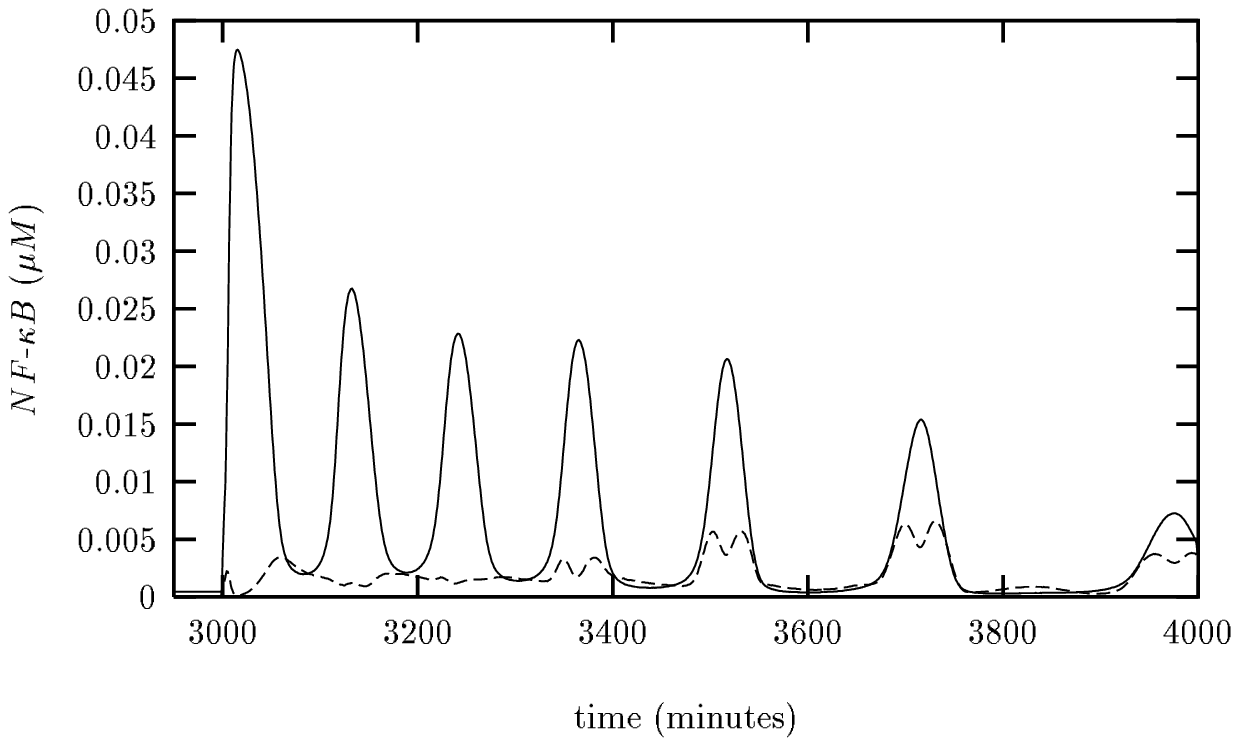}
\caption{}
\end{figure}

\begin{figure}[htbp]
\centering
\includegraphics{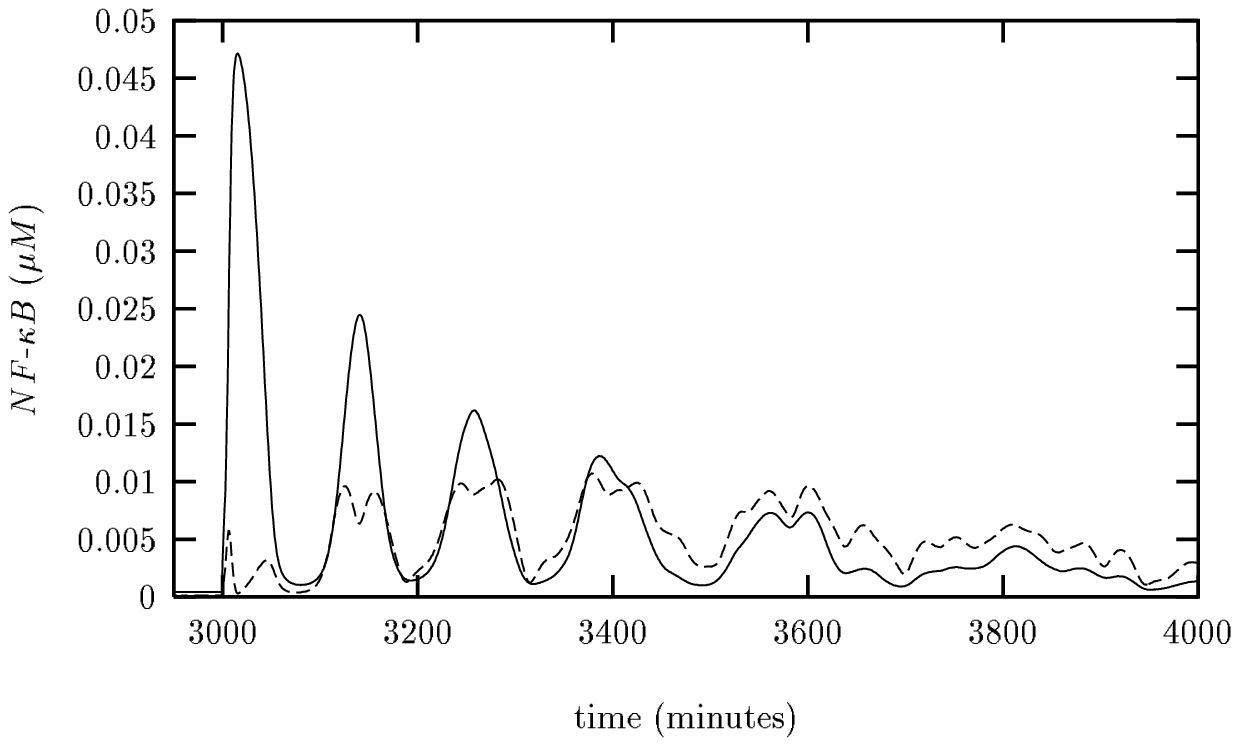}
\caption{}
\end{figure}

\begin{figure}[htbp]
\centering
\includegraphics{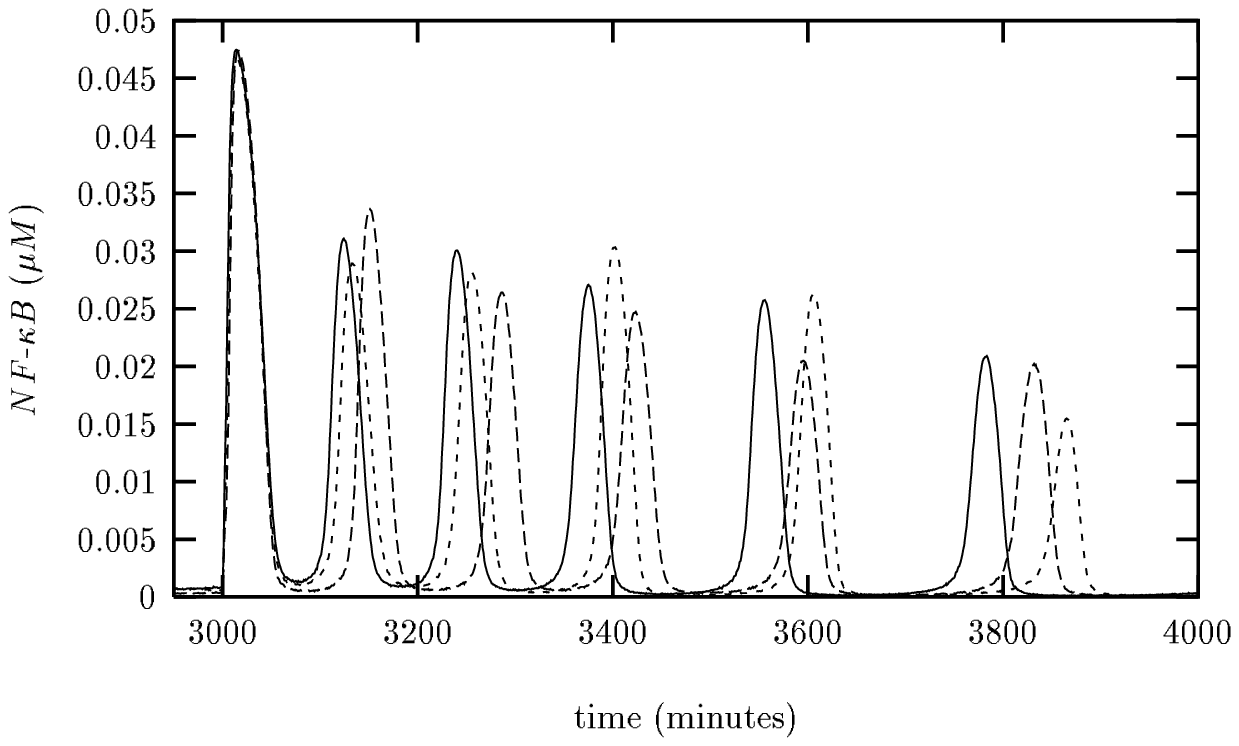}
\caption{}
\end{figure}

\begin{figure}[htbp]
\centering
\includegraphics{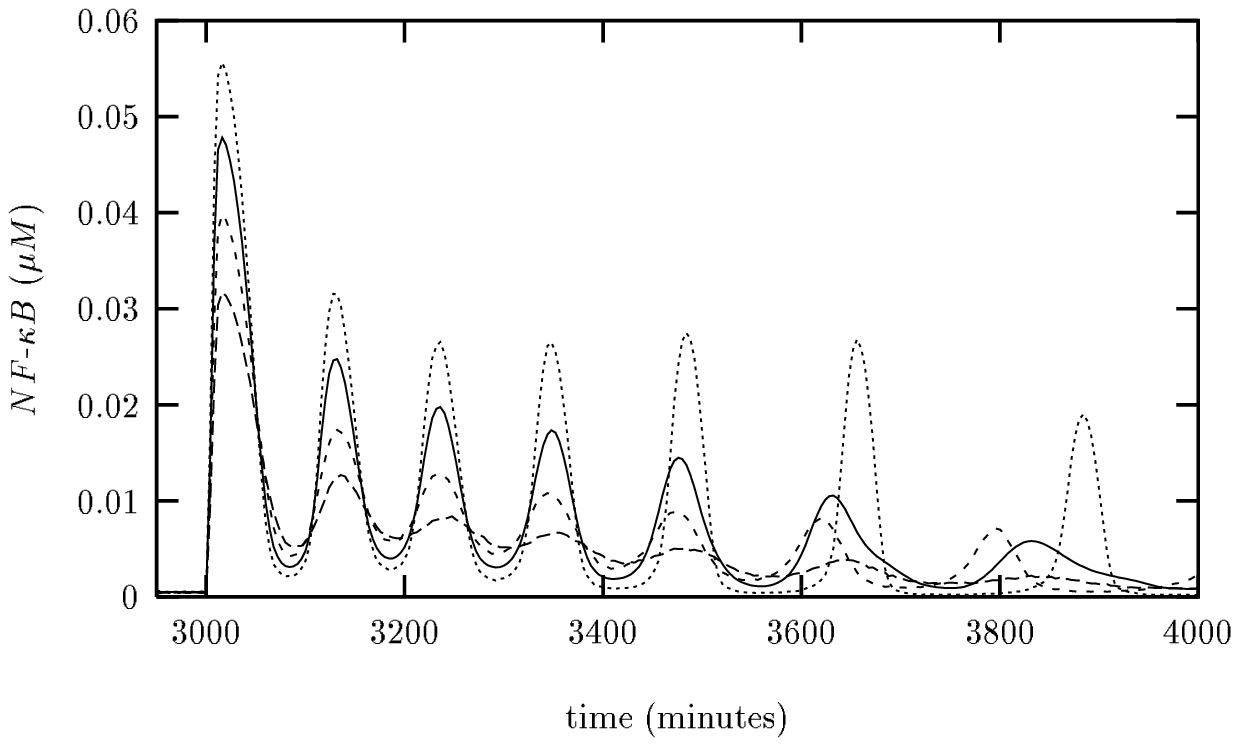}
\caption{}
\end{figure}

\begin{figure}[htbp]
\centering
\includegraphics{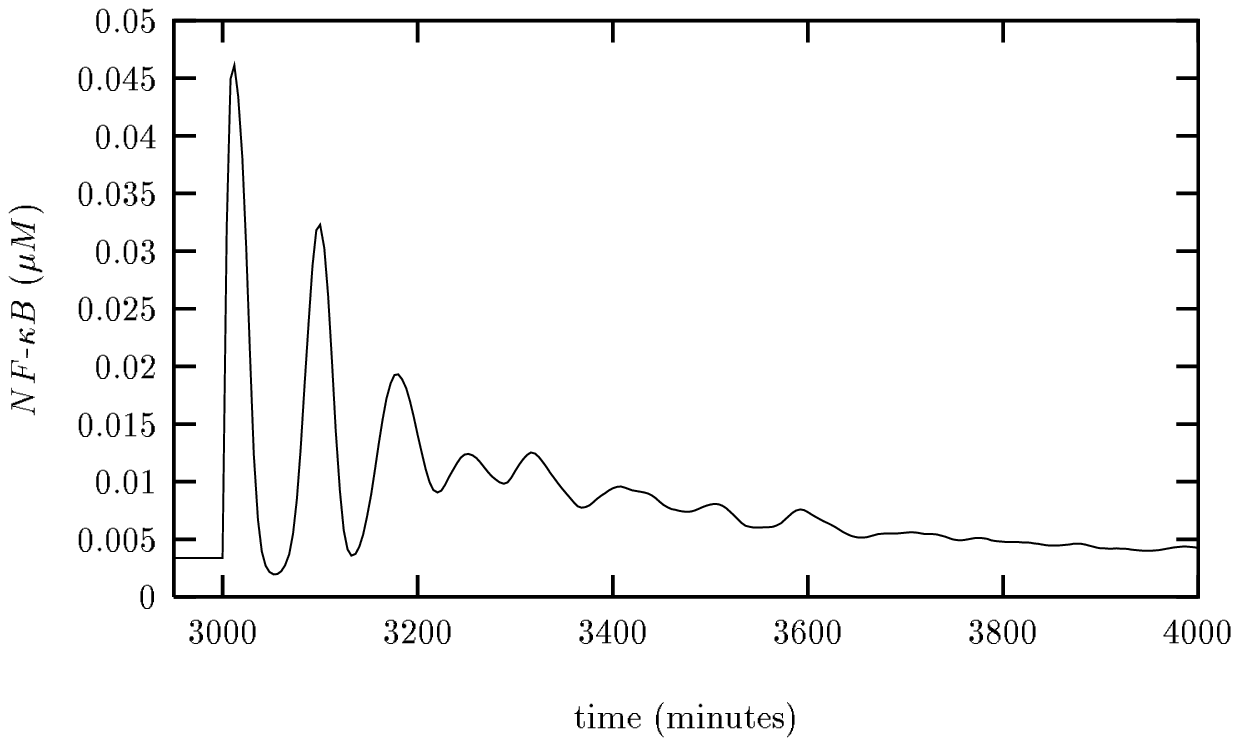}
\caption{}
\end{figure}

\begin{figure}[htbp]
\centering
\includegraphics{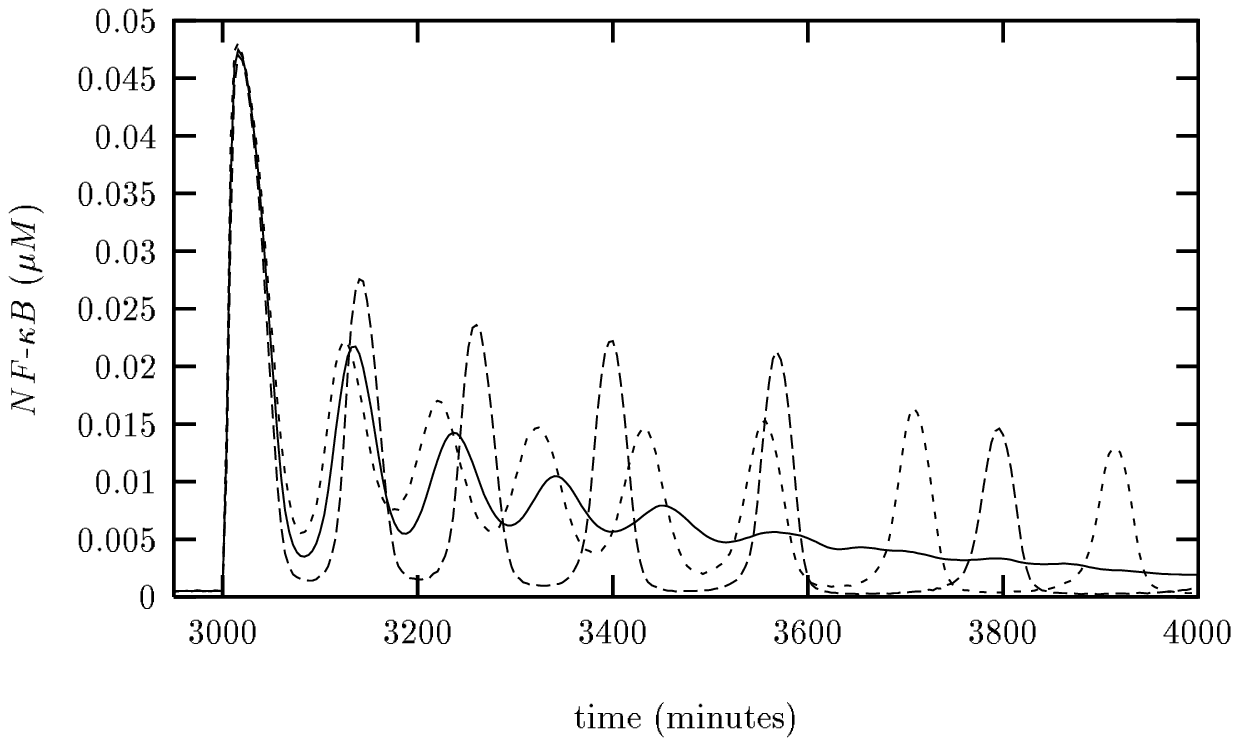}
\caption{}
\end{figure}

\begin{figure}[htbp]
\centering
\includegraphics{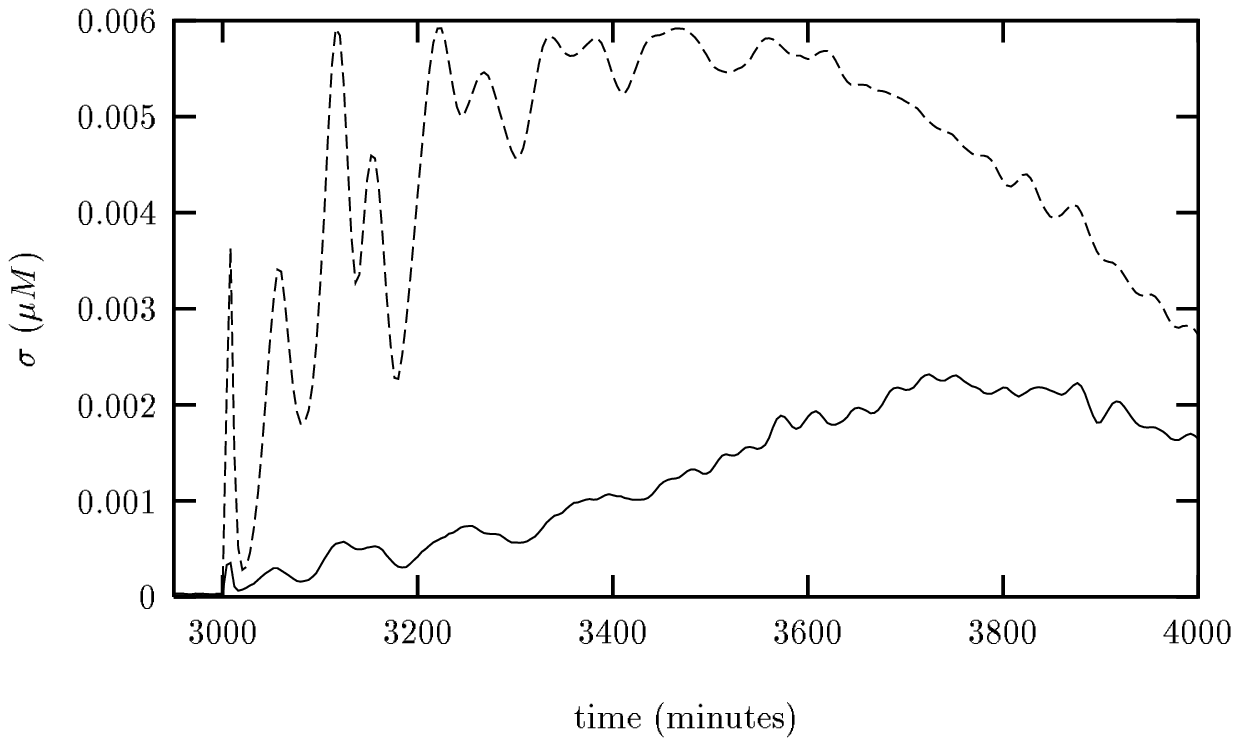}
\caption{}
\end{figure}

\end{document}